\documentclass[prb,showpacs]{revtex4}
\usepackage{amsmath}

\input{tcilatex}

\begin{document}

\title{The glass transition and the replica symmetry breaking in vortex
matter}
\author{$^{1,3}$Dingping Li and $^{2,3}$Baruch Rosenstein}
\affiliation{$^{1}$\textit{Department of Physics, Peking University, Beijing 100871,
China, P.R.C.}}
\affiliation{$^{2}$\textit{Department of Condensed Matter Physics, Weizmann Institute of
Science, Rehovot 76100, Israel}.}
\affiliation{$^{3}$\textit{National Center for Theoretical Sciences} \textit{and} \textit{%
Electrophysics Department, National Chiao Tung University,Hsinchu 30050,}
\textit{Taiwan, R.O.C}.}
\date{\today}

\begin{abstract}
We quantitatively describe the competition between the thermal fluctuations
and the disorder using the Ginzburg -- Landau approach. Flux line lattice in
type II superconductors undergoes a transition\textbf{\ }into three
\textquotedblleft disordered\textquotedblright\ phases: vortex liquid (not
pinned), homogeneous vortex glass (pinned) and crystalline Bragg glass
(pinned) due to both thermal fluctuations and random quenched disorder. We
show that disordered Ginzburg -- Landau model (valid not very far from $%
H_{c2}$) in which only the coefficient of a term quadratic in order
parameter $\psi $ is random first considered by Dorsey, Fisher and Huang
leads to a state with nonzero Edwards -- Anderson order parameter, but this
state is still replica symmetric. However when the coefficient of the
quartic term $\left\vert \psi \right\vert ^{4}$ \ in GL free energy also has
a random component, replica symmetry breaking effects appear. The location
of the glass transition line in 3D materials is determined and compared to
experiments. The line is clearly different from both the melting line and
the second peak line describing the translational and rotational symmetry
breaking at high and low temperatures respectively. The phase diagram is
therefore separated by two lines into four phases mentioned above.
\end{abstract}

\maketitle

\section{Introduction}

In any superconductor there are impurities either present naturally or
systematically produced using the proton or electron irradiation. The
inhomogeneities both on the microscopic and the mesoscopic scale greatly
affect thermodynamic and especially dynamic properties of type II
superconductors in magnetic field. The magnetic field penetrates the sample
in a form of Abrikosov vortices, which can be pinned by disorder. In
addition thermal fluctuations also greatly influence the vortex matter in
high $T_{c}$ superconductors, for example in some cases thermal fluctuations
will effectively reduce the effects of disorder. As a result the $H-T$ phase
diagram of the high $T_{c}$ superconductors is very complex due to the
competition between thermal fluctuations and disorder, and it is still far
from being reliably determined, even in the best studied superconductor, the
optimally doped $YBCO$ superconductor. \cite{Gammel} Difficulties are both
experimental and theoretical. Experimentally various phases with various
(frequently overlapping) names like liquid \cite{Nelson} (sometimes
differentiated into liquid I and liquid II \cite{Bouquet}), vortex solid,
Bragg glass \cite{GLD} (=pinned solid), vortex glass (=pinned
liquid=entangled solid, \cite{Ertas} the vortex slush \cite{Hu}), were
described. To differentiate various phases one should understand the nature
of the phase transitions between them. Although over the years the picture
has evolved with various critical and tricritical points appeared and
disappeared, several facts become increasingly clear.

1. The first order \cite{Zeldov95,Schilling} melting line seems to merge
with the "second magnetization peak" line forming the universal order -
disorder phase transition line. \cite{Zeldov96,Yeshurun} At the low
temperatures the location of this line strongly depends on disorder and
generally exhibits a positive slope (termed also the "inverse" melting \cite%
{Zeldov01}), while in the "melting" section it is dominated by thermal
fluctuations and has a large negative slope. The resulting maximum at which
the magnetization and the entropy jump vanish was interpreted either as a
tricritical point \cite{Nishizaki,Bouquet} or a Kauzmann point. \cite%
{Lidisorder} This universal "order - disorder" transition line (ODT), which
appeared first in the strongly layered superconductors ($BSCCO$ \cite%
{Zeldov96}) was extended to the moderately anisotropic superconductors ($%
LaSCCO$ \cite{Yeshurun}) and to the more isotropic ones like $YBCO$. \cite%
{Lidisorder,Grover} The symmetry characterization of the transition is
clear: spontaneous breaking of the translation and rotation symmetry.

2. The universal "order - disorder" line is different from the
"irreversibility line" or the "glass" transition (GT) line, which is a
continuous transition. \cite{Deligiannis,Taylor} The almost vertical glass
line clearly represents effects of disorder although the thermal
fluctuations affect the location of the transition. Experiments in $BSCCO$
\cite{Zeldov98} indicate that the line crosses the ODT line right at its
maximum, continues deep into the ordered (Bragg) phase. This proximity of
the glass line to the Kauzmann point is reasonable since both signal the
region of close competition of the disorder and the thermal fluctuations
effects. In more isotropic materials the data are more confusing. In $LaSCCO$
\cite{Forgan} the GT line is closer to the "melting" section of the ODT line
still crossing it. In YBCO we are not aware of a claim that the GT line
continuous into the ordered phase. Most of experiments \cite{Nishizaki}
indicate that the GT line terminates at the "tricritical point" in the
vicinity of the maximum of the ODT line. It is more difficult to
characterize the nature of the GT transition as a "symmetry breaking". The
common wisdom is that "replica" symmetry is broken in the glass (either via
"steps" or via "hierarchical" continuous process) as in the most of the spin
glasses theories. \cite{Dotsenko}

Theoretically the problem of the vortex matter subject to thermal
fluctuations or disorder has a long history. An obvious candidate to model
the disorder is the Ginzburg - Landau model in which coefficients have
random components. However this model is too complicated and simplifications
are required. The original idea of the vortex glass and the continuous glass
transition exhibiting the glass scaling of conductivity diverging in the
glass phase appeared early in the framework of the frustrated $XY$ model
(the gauge glass). \cite{Fisher,Natterman} In this approach one fixes the
amplitude of the order parameter retaining the magnetic field with random
component added to the vector potential. It \ was studied by the RG and the
variational methods and has been extensively simulated numerically. \cite%
{Olsson,Hu} In analogy to the theory of spin glass the replica symmetry is
broken when crossing the GT line. The model ran into several problems (see
Giamarchi and Bhattacharya in Ref. \onlinecite{GiamarchiBhattacharya} for a
review): for finite penetration depth $\lambda $ it has no transition \cite%
{BokilYoung} and there was a difficulty to explain sharp Bragg peaks
observed in the experiments at low magnetic fields. To address the last
problem another simplified model had been proven to be more convenient: the
elastic medium approach to a collection of interacting line-like objects
subject to both the pinning potential and the thermal bath Langevin force.
\cite{Otterlo,Reichhardt} The resulting theory was treated again using the
gaussian approximation \cite{Korshunov,GLD} and RG.\cite{Natterman} The
result was that in $2<D<4$ there is a transition to a glassy phase in which
the replica symmetry is broken following the \textquotedblleft hierarchical
pattern\textquotedblright\ (in $D=2$ the breaking is \textquotedblleft one
step\textquotedblright ). The problem of the very fast destruction of the
vortex lattice by disorder was solved with the vortex matter being in the
replica symmetry broken (RSB) phase and it was termed \textquotedblleft
Bragg glass\textquotedblright . \cite{GLD} \ It is possible to address the
problem of mesoscopic fluctuation using an approach in which one directly
simulates the interacting line-like objects subject to both the pinning
potential and the thermal bath Langevin force. \cite{Otterlo,Reichhardt} In
this context the generalized replicated density functional theory \cite%
{Menon02} was also applied resulting in one step RSB solution.\ Although the
above approximations to the disordered GL theory are very useful in more
\textquotedblleft fluctuating\textquotedblright\ superconductors like $BSCCO$%
, a problem arises with their application to $YBCO$ at temperature close $%
T_{c}$ (where most of the experiments mentioned above are done): vortices
are far from being line-like and even their cores significantly overlap. As
a consequence the behavior of the dense vortex matter is expected to be
different from that of a system of pointlike vortices and of the $XY$ model
although the elastic medium approximation might still be meaningful. \cite%
{Brandt}

To describe the non-pointlike vortices, one has to return to the GL model
and make a different simplification. One of the most developed schemes is
the lowest Landau level (LLL) approximation valid close to the $H_{c2}(T)$
line. \cite{Thouless} Such an attempt was made by Dorsey, Fisher and Huang
\cite{Dorsey} in the liquid phase using the dynamic approach \cite%
{Sompolinsky} and by Tesanovic and Herbut\ for columnar defects in layered
materials using supersymmetry. \cite{Herbut} It is the purpose of this paper
to study the glass transition using the replica formalism. We quantitatively
study the glass transition in the same model, with the disorder represented
by the random component of the coefficients of the GL free energy. The most
general hierarchical homogeneous (liquid) Ansatz \cite{Parisi} and its
stability are considered to obtain the glass transition line and to
determine the nature of the transition for various values of the disorder
strength of the GL coefficients. Then we place the glass line on the phase
diagram of YBCO and compare with experiments and other theories.

The paper is organized as follows. The general disordered GL model is
introduced in section II and the gaussian variational replica method is
presented in section III. Next we study in some detail the model either with
$\left\vert \psi \right\vert ^{2}$disorder in Section IV, with less details
the $\left\vert \psi \right\vert ^{4}$disorder in Section V and obtain the
phase transition lines in those two cases. In Section VI, the general model
containing both the $\left\vert \psi \right\vert ^{2}$disorder and the $%
\left\vert \psi \right\vert ^{4}$disorder is treated briefly. In Section
VII, we compare our results with the experimental data and conclude in
section VIII by summarizing our results.

\section{Disorder effects in the Ginzburg - Landau description of the type
II superconductor}

\subsection{Ginzburg - Landau free energy}

We start from the Gibbs energy of the ideal homogeneous sample (no disorder):%
\begin{equation}
G=\int dx^{3}\frac{{\hbar }^{2}}{2m_{{\tiny ||}}^{\ast }}\left\vert \partial
_{z}\psi \right\vert ^{2}+\frac{{\hbar }^{2}}{2m_{\perp }^{\ast }}\left\vert
\overrightarrow{D}\psi \right\vert ^{2}+a^{\prime }\psi ^{\ast }\psi +\frac{%
b^{\prime }}{2}(\psi ^{\ast }\psi )^{2}+\frac{(H-B)^{2}}{8\pi }.
\end{equation}%
Here $a^{\prime }=\alpha (T-T_{c})$ and $b^{\prime }$ are constant
parameters, $\overrightarrow{D}\equiv (-i{\hbar }\nabla +\frac{e^{\ast }}{c}%
\overrightarrow{A})$ is the covariant derivative, $\overrightarrow{A}$ is
the vector potential, the magnetic field $\overrightarrow{B}=\nabla \times
\overrightarrow{A}$, $H$ is the external magnetic field, $m_{\perp }^{\ast }$
and $m_{{\tiny ||}}^{\ast }$ are the effective masses in directions
perpendicular and parallel to the field respectively. Mesoscopic thermal
fluctuations are accounted for via Boltzmann weights%
\begin{equation}
Z=\int_{\psi ^{\ast },\psi }\exp \left\{ -\frac{G[\psi ^{\ast },\psi ]}{T}%
\right\}  \label{freeen}
\end{equation}

The model provides a good description of thermal fluctuations as long as $%
1-t-b<<1$, where $t=\frac{T}{T_{c}}$, $b=B/H_{c2}$ ($h=H/H_{c2}$), $%
H_{c2}=\Phi _{0}/(2\pi \xi ^{2})$ and $\xi $ is the coherence length. In
this case the higher order terms like $\left\vert \psi \right\vert ^{6}$ can
be omitted (detail notation can be found in Ref. \onlinecite{LiLiq}). The 3D
GL model describes materials with not too high anisotropy (for a recent
evidence of \ validity of this assumption in $YBCO$ see Ref. %
\onlinecite{Schilling2}). In strongly anisotropic materials, a \ model of
the Lawrence -- Doniach type is more appropriate. \cite{Blatter}

Within the GL approach the pointlike quenched disorder on the mesoscopic
scale is described by making \textbf{each} of the coefficients of the
mesoscopic GL energy a random value centered around a certain constant value
given in eq. (\ref{freeen}). For example effective masses can be disordered
\begin{eqnarray}
m_{\perp }^{\ast -1} &\rightarrow &m_{\perp }^{\ast -1}\left( 1+U(x)\right) ;
\\
\overline{\text{\ }U(x)U(y)} &=&P\delta (x-y).  \notag
\end{eqnarray}%
The parallel effective mass $m_{{\tiny ||}}^{\ast }$ might also have the
random component which we neglect (it is relatively small since $m_{{\tiny ||%
}}^{\ast }$ is typically very large), though it can be incorporated with no
additional difficulties. This type of disorder is sometimes called the $%
\delta l$ disorder since it originates in part from the inhomogeneity of the
electron mean free path $l$ in Gor'kov's derivation. From the BCS theory,
effective mass is $m^{\ast }=2m_{e}\left( 1+\frac{\pi ^{3}{\hbar }v_{F}}{%
168\zeta \left( 3\right) T_{c}l}\right) $ in the clean limit and $m^{\ast
}=2m_{e}\frac{7\zeta \left( 3\right) {\hbar v}_{F}}{2\pi ^{3}T_{c}l}$ in the
dirty limit. Relation to the notations of Ref. \onlinecite{Blatter} (chapter
II in this reference) is the following: $U$ is $-\delta m_{ab}/m_{ab}$ and $%
P=\gamma _{m}/m_{ab}^{2}$. Note however that, in addition to the random
distribution of $l,$ disorder in $v_{F}$ and $T_{c}$ (the density of states
and interaction strength) can also affect $m^{\ast }.$

\ The other two parameters in the GL equations are $\alpha =\frac{12\pi
^{2}T_{c}}{7\zeta \left( 3\right) \epsilon _{F}}$ and $b^{^{\prime }}=\frac{%
18\pi ^{2}}{7\zeta \left( 3\right) N\epsilon _{F}}\left( \frac{T_{c}}{%
\epsilon _{F}}\right) ^{2}$. The coefficient of the quadratic term is called
$\delta T$ disorder since it describes a local deviation of the critical
temperature. Introducing a random component in $\left\vert \psi \right\vert
^{2}$term:
\begin{equation}
a^{\prime }\rightarrow a^{\prime }(1+W(x));\text{ }\overline{\text{\ }%
W(x)W(y)}=R\delta (x-y).
\end{equation}%
In notations of Ref. \onlinecite{Blatter} the random field $W(x)=-\delta
_{a}/a,$ $R=\gamma _{a}/a^{2}.$ When thermal fluctuations of the vortex
degrees of freedom can be neglected, these two random fields would be
sufficient (they control the two relevant scales $\xi $ and $\lambda $). The
reason is that one can set the coefficient of the third term $|\psi |^{4}$
to a constant by rescaling. However in the presence of thermal fluctuations
the coefficient of $|\psi |^{4}$ also should be considered as having a
random component. It cannot be \textquotedblleft rescaled
out\textquotedblright\ since it affects the Boltzmann weights. We will see
later that at least within the lowest Landau level approximation this term
is crucial in inducing certain glassy properties of the vortex matter state.
We therefore introduce its disorder via%
\begin{equation}
b^{\prime }\rightarrow b^{\prime }(1+V(x));\text{ }\overline{\text{\ }%
V(x)V(y)}=Q\delta (x-y).
\end{equation}

In unconventional superconductors, even without disorder, the
phenomenological GL model has not been reliably derived \ microscopically.
The coefficients and their inhomogeneities therefore should be considered as
phenomenological parameters to be fitted to experiments. We assume that $U,$
$R$ and $Q$ have weak dependencies on field and temperature. The assumption
of the weak temperature and field dependence of the disorder strengths $U,$ $%
R$ and $Q$, as that of any parameter in the GL approach, should be derived
in principal from a microscopic theory assuming random chemical potential or
should be justified by fitting to experiments. For simplicity the white
noise distribution is considered
\begin{equation*}
p[U,W,V]=\exp \left[ -\int_{x}\frac{U(x)^{2}}{2P}+\frac{W(x)^{2}}{2R}+\frac{%
V(x)^{2}}{2Q}\right]
\end{equation*}%
for random components. The free energy of superconductor after averaging
over the disorder is
\begin{eqnarray}
\overline{F} &=&-\frac{T}{norm}\int_{U,W,V}p[U,W,V]\log \left[ \text{ }%
\int_{\psi }\text{exp}[-g[\psi ]-f_{dis}[U,W,V,\psi ]]\right] ;  \label{21}
\\
g &=&G/T;\text{ \ }f_{dis}[U,W,V,\psi ]=\frac{1}{T}\int_{x}\frac{{\hbar }^{2}%
}{2m_{\perp }^{\ast }}U(x)\left\vert \overrightarrow{D}\psi \right\vert
^{2}+a^{\prime }W(x)|\psi |^{2}+\frac{1}{2}V(x)|\psi |^{4},
\end{eqnarray}%
where$\ norm\equiv \int_{U,W,V}p[U,W,V]$ is a normalization factor.

To make the physical picture clear, we rescale the coordinates as $%
x\rightarrow \xi x,$ $y\rightarrow \xi y,$ $z\rightarrow \frac{\xi z}{\gamma
}$ with anisotropy parameter defined by $\gamma =\left( m_{{\tiny ||}}^{\ast
}/m_{\perp }^{\ast }\right) ^{1/2}$. The order parameter is scaled as $\psi
^{2}\rightarrow \frac{2\alpha T_{c}}{b^{\prime }}\psi ^{2}$. The
dimensionless free energy becomes simpler looking:

\begin{equation}
g[\psi ]=G/T=\frac{1}{\omega }\int_{x}\frac{{1}}{2}\left\vert \partial
_{z}\psi \right\vert ^{2}+\frac{{1}}{2}\left\vert \overrightarrow{D}\psi
\right\vert ^{2}+\frac{t-1}{2}|\psi |^{2}+\frac{1}{2}|\psi |^{4}+\frac{%
\kappa ^{2}\left( b-h\right) ^{2}}{4},
\end{equation}%
where $\omega =\sqrt{2Gi}\pi ^{2}t$. The Ginzburg number is $Gi\equiv 32%
\left[ \pi \lambda {^{2}}T_{c}\gamma {/(}\Phi {_{0}^{2}}\xi {)}\right] ^{2},$
{where} $\lambda $ is the magnetic penetration depth. The last term can be
ignored in calculating $\overline{F}$ as the $\kappa $ is very big in high $%
T_{c}$ superconductors and the last term is order of $\frac{1}{\kappa ^{2}}$%
. Similarly the random component and the distribution become:

\begin{eqnarray}
\text{\ }f_{dis}[U,W,V,\psi ] &=&\frac{1}{\omega }\int_{x}\left\{ -\frac{{1}%
}{2}U(x)\psi ^{\ast }D^{2}\psi +\frac{t-1}{2}W(x)\left\vert \psi \right\vert
^{2}+\frac{1}{2}V(x)\left\vert \psi \right\vert ^{4}\right\} \\
p[U,W,V] &=&\exp \left[ -\frac{\xi ^{3}}{\gamma }\int_{x}\left( \frac{%
U(x)^{2}}{2P}+\frac{W(x)^{2}}{2R}+\frac{V(x)^{2}}{2Q}\right) \right] .
\end{eqnarray}%
The model however is highly nontrivial even without disorder, and to make
progress, further approximation is needed.

\subsection{Lowest Landau level approximation}

The lowest Landau level (LLL) approximation \cite{Thouless} is based on
constraint $-D^{2}\psi =b\psi $. Over the years this model has been studied
by various methods, analytic and numerical. \cite{Ruggeri,Menon94,LiLiq} The
(effective) LLL model is applicable in a surprisingly wide range of fields
and temperatures determined by the condition that the relevant excitation
energy $\varepsilon $ is much smaller than the gap between Landau levels $%
2\hbar eB/(cm_{\perp })$.\cite{Lidisorder}

The free energy after further rescaling $x\rightarrow x/\sqrt{b}%
,y\rightarrow y/\sqrt{b},z\rightarrow z\left( \frac{2^{5/2}\pi }{b\omega }%
\right) ^{1/3},\psi ^{2}\rightarrow \left( \frac{2^{5/2}\pi }{b\omega }%
\right) ^{2/3}\psi ^{2}$, simplifies within the LLL approximation to:
\begin{equation}
f_{LLL}=\frac{1}{2^{5/2}\pi }\int d^{3}x\left[ \frac{1}{2}|\partial _{z}\psi
|^{2}+a_{T}|\psi |^{2}+\frac{1}{2}|\psi |^{4}\right] .  \label{GLscal}
\end{equation}%
Not surprisingly the number of independent constants in LLL is one less than
in the general model. This fact leads to the \textquotedblleft LLL
scaling\textquotedblright\ relations (of course the disorder terms will
break LLL scaling). As a result the simplified model without disorder has
just one parameter -- the (dimensionless) scaled temperature:
\begin{subequations}
\begin{equation}
a_{T}=-\left( \frac{2\pi }{b\omega }\right) ^{2/3}\left( 1-t-b\right) .
\label{aT3D}
\end{equation}%
The disorder term becomes:
\end{subequations}
\begin{equation}
f_{LLL}^{dis}=\frac{1}{2^{5/2}\pi }\int d^{3}x\left\{ \Omega (x)|\psi |^{2}+%
\frac{1}{2}V(x)|\psi |^{4}\right\} ,
\end{equation}%
in which only combination of $W$ and $U$ enters $\Omega (x)=\frac{1}{2}\left[
2\left( t-1\right) \left( \frac{2\pi }{b\omega }\right) ^{2/3}W(x)-bU(x)%
\right] $. Its distribution is still gaussian%
\begin{equation}
\overline{p}(\Omega ,V)=\exp \left[ -\int_{x}\frac{\Omega (x)^{2}}{%
2r^{\prime }}+\frac{V(x)^{2}}{2q}\right]
\end{equation}%
with two variances%
\begin{eqnarray}
r^{\prime } &=&\frac{\gamma }{4\sqrt{2}\xi ^{3}}\left\{ \frac{8\pi }{\omega }%
\left( 1-t\right) ^{2}R+\left( \frac{b^{2}\omega }{2\pi }\right)
^{1/3}b^{2}P\right\} \\
\text{\ \ \ }q^{\prime } &=&\frac{\gamma }{\sqrt{2}\xi ^{3}}\left( \frac{%
b^{2}\omega }{2\pi }\right) ^{1/3}Q.  \notag
\end{eqnarray}%
To treat both the thermal fluctuations and disorder we will use the replica
method to integrate over impurity distribution followed by gaussian
approximation.

\bigskip

\section{Replica trick and gaussian approximation}

\subsection{Replica trick}

We will use the replica trick to evaluate the disorder averages. The replica
method is widely used to study disordered electrons in the theory of spin
glasses, \cite{Dotsenko} disordered metals and was applied to vortex matter
in the London limit. \cite{Giamarchi,Korshunov} Applying a simple
mathematical identity to the disorder average of the free energy one
obtains:
\begin{equation}
\overline{F}=\overline{-T\underset{n\rightarrow 0}{\lim }\frac{1}{n}(Z^{n}-1)%
}.  \label{23}
\end{equation}%
The averages\ of $Z^{n}$ is the statistical sum over $n$ identical "replica"
fields $\psi _{a}$ , $a=1,...,n$:
\begin{equation}
\overline{Z^{n}}=\frac{1}{norm}\int_{\Omega ,V}p[\Omega
,V]\prod_{a}\int_{\psi _{a}}\text{exp}\left\{ -f[\psi _{a}]-f_{dis}[\Omega
,V,\psi _{a}]\right\} .
\end{equation}%
The integral over the disorder potential is gaussian and results in:

\begin{eqnarray}
\overline{Z^{n}} &=&\int_{\psi _{a}}\exp \left[ -\sum_{a}f(\psi _{a})+\frac{1%
}{2\left( 2^{5/2}\pi \right) ^{2}}\sum_{a,b}f_{ab}\right] \\
f_{ab} &=&r^{\prime }\left\vert \psi _{a}\right\vert ^{2}\left\vert \psi
_{b}\right\vert ^{2}+\frac{q^{\prime }}{4}(\psi _{a}^{\ast }\psi
_{a})^{2}(\psi _{b}^{\ast }\psi _{b})^{2}.  \notag
\end{eqnarray}%
This model is a type of scalar field theory and the simplest nonperturbative
scheme commonly used to treat such a model is gaussian approximation. Its
validity and precision can be checked only by calculating corrections.

\subsection{\protect\bigskip Gaussian approximation}

We have assumed that the order parameter is constrained to the LLL and
therefore can be expanded in a basis of the standard LLL eigenfunctions in
Landau gauge:
\begin{equation}
\psi _{a}(x)=norm\int_{k_{z},k}e^{i\left( zk_{z}+xk\right) }\exp \left\{ -%
\frac{1}{2}(y+k)^{2}\right\} \widetilde{\psi }_{a}(k).
\end{equation}%
We now apply the gaussian approximation which has been used in disorder in
the elastic medium approach, \cite{Korshunov,Giamarchi} following its use in
polymer physics. \cite{Mezard} The gaussian approximation was applied to the
vortex liquid within the GL approach in \onlinecite{Thouless,Ruggeri} The
gaussian effective free energy is expressed via variational parameter \cite%
{Mezard,LiLiq} $\mu _{ab}$ which in the present case is a matrix in the
replica space. The correlator is parametrized as%
\begin{equation}
\left\langle \psi _{a}^{\ast }(k,k_{z})\psi _{b}(-k,-k_{z})\right\rangle
=G_{ab}(k_{z})=\frac{2^{5/2}\pi }{\frac{k_{z}^{2}}{2}\delta _{ab}+\mu
_{ab}^{2}}
\end{equation}%
The bubble integral appearing in the free energy is very simple:%
\begin{equation*}
\left\langle \psi _{a}^{\ast }(x,y,z)\psi _{b}(x,y,z)\right\rangle =\frac{%
\sqrt{2}}{\pi }\int_{k_{z}}\frac{1}{\frac{k_{z}^{2}}{2}\delta _{ab}+\mu
_{ab}^{2}}=2\mu _{ab}^{-1}\equiv 2m_{ab}.
\end{equation*}%
As a result the gaussian effective free energy can be written in a form:%
\begin{eqnarray}
n\text{ }f_{eff} &=&\sum_{a}\left\{ \frac{2^{5/2}\pi }{\left( 2\pi \right)
^{3}}\int_{k_{z}}\left[ LogG^{-1}(k_{z})+\left( \frac{k_{z}^{2}}{2}%
+a_{T}\right) G(k_{z})-I\right] _{aa}+4\left( m_{aa}\right) ^{2}\right\}
\notag \\
&&-\sum_{a,b}\left\{ \frac{1}{2^{3/2}\pi }r^{\prime }\left\vert
m_{ab}\right\vert ^{2}+\frac{\sqrt{2}}{\pi }q^{\prime }\left( \left\vert
m_{ab}\right\vert ^{4}+4m_{aa}m_{bb}\left\vert m_{ab}\right\vert ^{2}\right)
\right\} \\
&=&2\sum_{a}\left\{ \mu _{aa}+a_{T}m_{aa}+2\left( m_{aa}\right) ^{2}\right\}
\notag \\
&&-2\sum_{a,b}\left\{ r\left\vert m_{ab}\right\vert ^{2}+q\left( \frac{1}{4}%
\left\vert m_{ab}\right\vert ^{4}+m_{aa}m_{bb}\left\vert m_{ab}\right\vert
^{2}\right) \right\} ,  \notag
\end{eqnarray}%
where we discarded an (ultraviolet divergent) constant and renormalization
of $a_{T}$ and rescaled the disorder strength: $r=\frac{1}{2^{5/2}\pi }%
r^{\prime },$ \ $q=\frac{2^{5/2}}{\pi }q^{\prime }.$

We start with a simple case in which only the $\left\vert \psi \right\vert
^{2}$ type of disorder is present. More precisely we take $q=0$ and return
to the general case in section IV. This model has been already discussed
using different method (the Sompolinsky dynamic approach) in the unpinned
phase in Ref. \onlinecite{Dorsey}.

\section{Nonzero Edwards - Anderson order parameter and absence of the
replica symmetry breaking when only the $\left\vert \protect\psi \right\vert
^{2}$ disorder is present}

\subsection{Hierarchical matrices and impossibility of the continuous
replica symmetry breaking}

In this section we neglect the $\left\vert \psi \right\vert ^{4}$ disorder
term. It is convenient to introduce real (not necessarily symmetric) matrix $%
Q_{ab},$ which is in one to one linear correspondence with Hermitian
(generally complex) matrix $m_{ab}$ via%
\begin{equation}
Q_{ab}=\text{re}[m_{ab}]+\text{im}[m_{ab}].  \label{Qdef}
\end{equation}%
Unlike $m_{ab}$, all the matrix elements of $Q_{ab}$ are independent. In
terms of this matrix the free energy can be written as

\begin{equation}
\frac{n}{2}f_{eff}=\sum_{a}\left\{ \left( m^{-1}\right)
_{aa}+a_{T}Q_{aa}+2\left( Q_{aa}\right) ^{2}\right\} -r\sum_{a,b}Q_{ab}^{2}.
\label{fR}
\end{equation}%
Taking derivative with respect to $Q_{ab}$ gives the saddle point equation
for this matrix element:

\begin{equation}
\frac{n}{2}\frac{\delta f}{\delta Q_{ab}}=-\frac{1}{2}\left[ \left(
1-i\right) \left( m^{-2}\right) _{ab}+c.c.\right] +a_{T}\delta
_{ab}+4Q_{aa}\delta _{ab}-2rQ_{ab}=0.  \label{mateqR}
\end{equation}%
Since the electric charge (or the superconducting phase) $U(1)$ symmetry is
assumed, we consider only solutions with real $m_{ab}.$ In this case $%
m_{ab}=Q_{ab}$ is a symmetric real matrix. General hierarchical matrices $m$
are parametrized using the diagonal elements $\widetilde{m}$ and the
Parisi's (monotonically increasing) function $m_{x}$ specifying the off
diagonal elements with $0<x<1$. \cite{Mezard} Physically different $x$
represent time scales in the glass phase. In particular the Edwards -
Anderson (EA) order parameter is $m_{x=1}=M>0$.

A nonzero value for this order parameter signals that the annealed and the
quenched averages are different. The dynamic properties of such phase are
generally quite different from those of the nonglassy $M=0$\ phase. In
particular it is expected to exhibit infinite conductivity. \cite%
{Fisher,Dorsey} We will refer to this phase as the "ergodic pinned liquid"
(EPL) distinguished from the "nonergodic pinned liquid" (NPL) in which, in
addition, the ergodicity is broken.

However in the present model RSB does not occur. In terms of $\ $Parisi
parameter\ \ $\widetilde{m}$ and $m_{x}$ the matrix equation eq.(\ref{mateqR}%
) takes a form:

\begin{equation}
-\widetilde{m^{-2}}+a_{T}+\left( 4-2r\right) \widetilde{m}=0  \label{spR}
\end{equation}%
\begin{equation*}
\left( m^{-2}\right) _{x}+2rm_{x}=0.
\end{equation*}%
Dynamically if $m_{x}$ is a constant, \ pinning does not results in the
multitude of time scales. Certain time scale sensitive phenomena like
various memory effects \cite{Andrei} and the responses to \textquotedblleft
shaking\textquotedblright\ \cite{Zeldov01} are expected to be different from
the case when $m_{x}$ takes multiple values. If $m_{x}$ takes a finite
different number of $n$ values, we call $n-1$ step RSB. On the other hand,
if $m_{x}$ is continuous, the continuous replica symmetry breaking (RSB)
occurs.

In order to show that $m_{x}$ is a constant, it is convenient to rewrite the
second equation via the matrix $\mu ,$ the matrix inverse to $m$:

\begin{equation}
\left( \mu ^{2}\right) _{x}+2r(\mu ^{-1})_{x}=0.
\end{equation}%
Differentiating this equation with respect to $x$ one obtains;

\begin{equation}
2\left[ \left\{ \mu \right\} _{x}-r\left( \left\{ \mu \right\} _{x}\right)
^{-2}\right] x\frac{d\mu _{x}}{dx}=0,  \label{proof1}
\end{equation}%
where we used a set of standard notations in the spin glass theory: \cite%
{Mezard}
\begin{equation}
\left\{ \mu \right\} _{x}\equiv \widetilde{\mu }-\left\langle \mu
_{x}\right\rangle -[\mu ]_{x};\text{ \ \ }\left\langle \mu _{x}\right\rangle
\equiv \int_{0}^{1}dx\mu _{x};\text{ \ \ }[\mu ]_{x}=\int_{0}^{x}dy\left(
\mu _{x}-\mu _{y}\right) .
\end{equation}%
If one is interested in a continuous monotonic part $\frac{d\mu _{x}}{dx}%
\neq 0,$ the only solution of eq.(\ref{proof1}) is%
\begin{equation}
\left\{ \mu \right\} _{x}=r^{1/3}
\end{equation}%
Differentiating this again and dropping the nonzero derivative $\frac{d\mu
_{x}}{dx}$ again, one further gets a contradiction: $\frac{d\mu _{x}}{dx}=0$
. This proves that there are no such monotonically increasing continuous
segments. One can therefore generally have either the replica symmetric
solutions, namely $m_{x}=M$ or look for a several steplike RSB solutions.
\cite{Dotsenko} We can show that the constant $m_{x}$ solution is stable.
Therefore, if a steplike RSB solution exists, it might be only an additional
local minimum. We explicitly looked for a one step solution and found that
there is none.

\subsection{Two replica symmetric solutions\ and the third order transition
between them}

\subsubsection{The unpinned liquid and the \textquotedblleft ergodic
glass\textquotedblright\ replica symmetric solutions}

\bigskip

Restricting to RS solutions, $m_{x}=M,$ the saddle point equations eq.(\ref%
{spR}) simplify:%
\begin{eqnarray}
-\varepsilon ^{-2}+\left( a_{T}+4\widetilde{m}\right) -2r\varepsilon &=&0;
\label{Req} \\
M\left( \varepsilon ^{-3}-r\right) &=&0,  \notag
\end{eqnarray}%
where $\varepsilon \equiv \widetilde{m}-M$. Energy of such a solution is
given by
\begin{equation}
\frac{f_{eff}}{2}=2\varepsilon ^{-1}-2-\varepsilon ^{-2}M+2a_{T}\widetilde{m}%
+4\widetilde{m}^{2}-2r\left( \varepsilon ^{2}+2\varepsilon M\right) .
\end{equation}%
The second equation eq.(\ref{Req}) has a replica index independent
(diagonal) solution $M=0$. In addition there is a non diagonal one. It turns
out that there is a third order transition between them.

For the diagonal solution $\varepsilon =\widetilde{m}$ and the first
equation is just a cubic equation:

\begin{equation}
-\widetilde{m}^{-2}+\left( a_{T}+4\widetilde{m}\right) -2r\widetilde{m}=0.
\label{diageq}
\end{equation}%
For the non diagonal solution the second equation gives $\varepsilon
=r^{1/3} $, which, when plugged into the first equation, gives:%
\begin{equation}
\widetilde{m}=\frac{1}{4}\left( 3r^{2/3}-a_{T}\right) ;\text{ \ }M=\frac{1}{4%
}\left( 3r^{2/3}-a_{T}\right) -r^{-1/3}.
\end{equation}%
The matrix $m$ therefore is
\begin{equation}
m_{ab}=r^{-1/3}\delta _{ab}+M,
\end{equation}%
which results in the following value of the free energy: $f=6r^{1/3}-\frac{1%
}{4}\left( 3r^{2/3}-a_{T}\right) ^{2}$.

The two solutions coincide for%
\begin{equation}
a_{T}=r^{-1/3}\left( 3r-4\right) .  \label{dynlineq}
\end{equation}%
Since in addition to the energy, the first and second derivatives of the
energy, $\frac{df}{da_{T}}=2r^{-1/3}$ and $\frac{d^{2}f}{da_{T}^{2}}=-\frac{1%
}{2}$ respectively, coincide (the fourth derivatives are different though),
the transition is a third order one.

\subsubsection{Stability domains of the two solutions}

In order to prove that a solution is stable beyond the set of replica
symmetric matrices $m$ one has to calculate the second derivative of free
energy (called Hessian in Refs. \onlinecite{Alemeida,Dotsenko}) with respect
to arbitrary real matrix $Q_{ab}$ defined in eq.(\ref{Qdef}):

\begin{eqnarray}
H_{(ab)(cd)} &\equiv &\frac{n}{2}\frac{\delta ^{2}f_{eff}}{\delta
Q_{ab}\delta Q_{cd}} \\
&=&\frac{1}{2}\left[ \left( m^{-2}\right) _{ac}\left( m^{-1}\right)
_{db}-i\left( m^{-2}\right) _{ad}\left( m^{-1}\right) _{cb}\right] +  \notag
\\
&&\frac{1}{2}\left[ \left( m^{-1}\right) _{ac}\left( m^{-2}\right)
_{db}-i\left( m^{-1}\right) _{ad}\left( m^{-2}\right) _{cb}\right] +cc
\notag \\
&&+4\delta _{ac}\delta _{bd}\delta _{ab}-2r\delta _{ac}\delta _{bd}.  \notag
\end{eqnarray}%
We will use a simplified notation for\ the product of the Kronecker delta
functions with more than two indices: $\delta _{ac}\delta _{bd}\delta
_{ab}\equiv \delta _{abcd}.$ For the diagonal solution the Hessian is a very
simple operator on the space of real symmetric matrices:

\begin{equation}
H_{(ab)(cd)}=c_{I}I_{abcd}+c_{J}J_{abcd},
\end{equation}%
where the operators $I$ (the identity in this space) and $J$ are defined as
\begin{equation}
I\equiv \delta _{ac}\delta _{bd};\text{ \ \ \ \ }J=\delta _{abcd}
\end{equation}%
and their coefficients in the diagonal phase are:

\begin{equation}
c_{I}=2\left( \widetilde{m}^{-3}-r\right) ,\text{ \ \ }c_{J}=4
\end{equation}%
with $\widetilde{m}$ being a solution of eq.(\ref{diageq}). The
corresponding eigenvectors in the space of symmetric matrices are $%
v_{(cd)}\equiv A\delta _{cd}+B.$ To find eigenvalues $\lambda $ of $H$ we
apply the Hessian on $V.$ The result is (dropping terms vanishing in the
limit $n\rightarrow 0$):%
\begin{equation}
H_{(ab)(cd)}v_{cd}=A\left( c_{I}+c_{J}\right) \delta _{ab}+B\left(
c_{I}+c_{J}\delta _{ab}\right) =\lambda \left( A\delta _{ab}+B\right)
\end{equation}%
There two eigenvalues: $\lambda ^{(1)}=c_{I}$ and $\lambda
^{(2)}=c_{I}+c_{J} $. Since $c_{J}=4>0,$ the sufficient condition for
stability is:
\begin{equation}
c_{I}=2\left( \widetilde{m}^{-3}-r\right) >0.
\end{equation}%
It is satisfied everywhere below the transition line of eq.(\ref{dynlineq}),
see Fig 1.( $a_{T},r$ phase diagram). The analysis of stability of the non
diagonal solution is slightly more complicated. The Hessian for the non
diagonal solution is:

\begin{equation}
H_{(ab)(cd)}=c_{V}V+c_{U}U+c_{J}J,
\end{equation}%
where new operators are%
\begin{equation}
V_{(ab)(cd)}=\delta _{ac}+\delta _{bd};U_{(ab)(cd)}=1
\end{equation}%
and coefficients are

\begin{equation}
c_{V}=-3Mr^{2/3};\text{ \ \ }c_{U}=4M^{2}r^{1/3};c_{J}=4
\end{equation}%
In the present case, one obtains three different eigenvalues, \cite%
{Alemeida,Dotsenko} $\lambda ^{(1.2)}=2\left( 1\pm \sqrt{1-4Mr^{2/3}}\right)
$ and $\lambda ^{(3)}=0.$ Note that the eigenvalue of Hessian on the
antisymmetric matrices are degenerate with eigenvalue $\lambda ^{(1)}$ in
this case (we will come back later on this eigenvalue). For $M<0$ the
solution is unstable due to negative $\lambda ^{(2)}.$ For $M>0,$ both
eigenvalues are positive and the solution is stable. The line $M=0$
coincides with the third order transition line, hence the non diagonal
solution is stable when the diagonal is unstable and \textit{vise versa}. We
conclude that there is no glass state in the vortex liquid without the $%
\left\vert \psi \right\vert ^{4}$ disorder term. The transition does not
correspond to RSB. Despite this in the phase with nonzero EA (NEA) $M$ order
parameter there are Goldstone bosons corresponding to $\lambda ^{(3)}$ in
the replica limit of $n\rightarrow 0$. The criticality and the zero modes
due to disorder (pinning) in this phase might lead to great variety of
interesting phenomena in statics and dynamics. These have not been explored
yet. However, as we show in the next section, the random component of the
quartic term changes the character of the transition line: the replica
symmetry is broken on the one side of the line. For simplicity in the next
section, we consider first a case with a random component of $\left\vert
\psi \right\vert ^{4}$ and no random component of $\left\vert \psi
\right\vert ^{2}$, and return to the general case in section V.

\section{\protect\bigskip The glass transition for the $\left\vert \protect%
\psi \right\vert ^{4}$ disorder}

\subsection{\protect\bigskip Continuous replica symmetry breaking solutions}

In this section we neglect the $r\left\vert \psi \right\vert ^{2}$ term
disorder. Although it is always present, as we have seen in the previous
section, at least within the gaussian approximation, it does not cause
replica symmetry breaking. Therefore one expects that although it certainly
influences properties of the vortex matter, for example, the melting
transition line to lower fields and temperatures, \cite{Lidisorder} it's
role in qualitative understanding of RSB effects is minor. The only other
disordered term within the LLL approximation considered in this paper is the
$\left\vert \psi \right\vert ^{4}$ disorder term. As was discussed in
section II, at least within the BCS theory, it is expected to be smaller
than the $\left\vert \psi \right\vert ^{2}$ disorder, $q\ll r$. Even it
could be very small, however, as we show here, it leads to qualitatively new
phenomena in vortex matter. The $r=0$ free energy after integration over $%
k_{z}$ becomes:

\begin{equation}
\frac{n}{2}f_{eff}=\sum_{a}\left\{ \left( m^{-1}\right)
_{aa}+a_{T}m_{aa}+2\left( m_{aa}\right) ^{2}\right\} -q\sum_{a,b}\left(
\frac{1}{4}\left\vert m_{ab}\right\vert ^{4}+m_{aa}m_{bb}\left\vert
m_{ab}\right\vert ^{2}\right)  \label{fQ}
\end{equation}%
In terms of the real matrix $Q_{ab}$ defined in eq.(\ref{Qdef}) the free
energy can be written as

\begin{eqnarray}
\frac{n}{2}f_{eff} &=&\sum_{a}\left\{ \left( m^{-1}\right)
_{aa}+a_{T}Q_{aa}+2\left( Q_{aa}\right) ^{2}\right\} \\
&&-q\sum_{a,b}\left( \frac{1}{8}Q_{ab}^{4}+\frac{1}{8}%
Q_{ab}^{2}Q_{ba}^{2}+Q_{aa}Q_{bb}Q_{ab}^{2}\right)
\end{eqnarray}%
Taking a derivative with respect to $Q_{ab}$ gives the saddle point equation
for this matrix:

\begin{equation}
\frac{n}{2}\frac{\delta f}{\delta Q_{ab}}=\left[
\begin{array}{c}
-\frac{1}{2}\left[ \left( 1-i\right) \left( m^{-2}\right) _{ab}+cc\right]
+a_{T}\delta _{ab}+4Q_{aa}\delta _{ab} \\
-q\left( \frac{1}{2}Q_{ab}^{3}+\frac{1}{2}%
Q_{ab}Q_{ba}^{2}+2Q_{ab}Q_{aa}Q_{bb}+\delta _{ab}\sum_{e}Q_{ee}\left(
Q_{ae}^{2}+Q_{ea}^{2}\right) \right)%
\end{array}%
\right] =0
\end{equation}%
Using the hierarchical symmetric matrix parametrization of its symmetric
part (the antisymmetric will not be important for most of our purposes), it
takes a form

\begin{equation}
-\widetilde{m^{-2}}+a_{T}+4\widetilde{m}-q\left( 3\widetilde{m}^{3}+2%
\widetilde{m^{2}}\widetilde{m}\right) =0  \label{tildQ}
\end{equation}%
\begin{equation}
\left( m^{-2}\right) _{x}+q\left( m_{x}^{3}+2\widetilde{m}^{2}m_{x}\right)
=0.  \label{xQ}
\end{equation}

As in the previous section, it is convenient to rewrite the second equation
in terms of $\mu ,$ the inverse matrix of $m$:

\begin{equation}
\left( \mu ^{2}\right) _{x}+q\left[ \left( (\mu ^{-1})_{x}\right) ^{3}+2%
\widetilde{m}^{2}(\mu ^{-1})_{x}\right] =0.
\end{equation}%
Differentiation of this equation with respect to $x$ leads to:

\begin{equation}
\left\{ 2\left\{ \mu \right\} _{x}-q\left[ 3\left( (\mu ^{-1})_{x}\right)
^{2}+2\widetilde{m}^{2}\right] \left( \left\{ \mu \right\} _{x}\right)
^{-2}\right\} x\frac{d\mu _{x}}{dx}=0.  \label{eqq}
\end{equation}%
For a continuous segment $\frac{d\mu _{x}}{dx}\neq 0$ one solves eq.(\ref%
{eqq}) for $(\mu ^{-1})_{x}$ in terms of $\left\{ \mu \right\} _{x}$ getting
now a more complicated result:

\begin{equation}
(\mu ^{-1})_{x}=\sqrt{\frac{2}{3}\left[ q^{-1}\left( \left\{ \mu \right\}
_{x}\right) ^{3}-\widetilde{m}^{2}\right] }
\end{equation}%
Differentiating this equation with respect to $x$ again one obtains:

\begin{equation}
\frac{1}{\left( \left\{ \mu \right\} _{x}\right) ^{2}}=\sqrt{\frac{2}{3q%
\left[ \left( \left\{ \mu \right\} _{x}\right) ^{3}-q\widetilde{m}^{2}\right]
}}\left( \left\{ \mu \right\} _{x}\right) ^{2}x.
\end{equation}%
Instead of solving this for $\left\{ \mu \right\} _{x}$ we present $x$ as
function of $\left\{ \mu \right\} _{x}$:

\begin{equation}
x=\sqrt{\frac{3q\left[ \left( \left\{ \mu \right\} _{x}\right) ^{3}-q%
\widetilde{m}^{2}\right] }{2\left( \left\{ \mu \right\} _{x}\right) ^{8}}}.
\label{xq}
\end{equation}

Thus the solution will be given by eq.(\ref{xq}) in the segment if $\frac{%
d\mu _{x}}{dx}\neq 0$ and constant $\mu _{x}$ in the other segments. In
principle this would allow for a numerical solution. One could actually
solve the equation near the transition line using the method in Ref. %
\onlinecite{Parisi}. The situation is completely different compared to that
of the $\left\vert \psi \right\vert ^{2}$ disorder. In the present case a
stable RSB solution exists. We will turn first however to the replica
symmetric solutions and determine their region of stability. In the unstable
region of the replica symmetric solutions, the RSB solution of eq.(\ref{xq})
will be the relevant one.

\subsection{Two replica symmetric solutions\protect\bigskip\ }

\subsubsection{Solutions}

Here we briefly repeat the steps leading to the RS solutions for the $%
\left\vert \psi \right\vert ^{2}$ disorder omitting details. The saddle
point equations eq.(\ref{spR}) for the RS matrices $m_{x}=M$ are:

\begin{equation}
-\varepsilon ^{-2}+\left( a_{T}+4\widetilde{m}\right) -q\left[ 5\widetilde{m}%
^{3}-M^{3}-2M^{2}\widetilde{m}-2\widetilde{m}^{2}M\right] =0  \label{eq1Q}
\end{equation}

\begin{equation}
M\left[ 2\varepsilon ^{-3}-q\left( M^{2}+2\widetilde{m}^{2}\right) \right] =0
\label{eq2Q}
\end{equation}%
where $\varepsilon \equiv \widetilde{m}-M$. Energy of such a solution is
given by
\begin{equation}
\frac{f}{2}=\varepsilon ^{-1}-\varepsilon ^{-2}M+a_{T}\widetilde{m}+2%
\widetilde{m}^{2}-\frac{q}{4}\left( 5\widetilde{m}^{2}-M^{2}\right) \left(
\widetilde{m}^{2}+M^{2}\right) .
\end{equation}%
For the diagonal solution $M=0,$ $\varepsilon =\widetilde{m}$ and the first
equation takes a form:

\begin{equation}
-\widetilde{m}^{-2}+a_{T}+4\widetilde{m}-5q\widetilde{m}^{3}=0.
\end{equation}%
The non diagonal solution in the present case is more complicated, but the
condition determining the transition line between the two (equivalently the
appearance of the nonvanishing EA order parameter) is still very simple: $%
\varepsilon =q^{-1/5}$ as $M=0$ on the line. Along the line the scaled
temperature is:%
\begin{equation}
a_{T}^{d}=2(3q^{2/5}-2q^{-1/5}).  \label{lowerl}
\end{equation}%
It is still the third order transition line similar to the $\left\vert \psi
\right\vert ^{2}$ disorder case and one has zero modes in NEA sector, while
no such modes exist in the $M=0$ phase. The stability analysis with respect
to configurations which are replica symmetric however gives completely
different results compared to that of the $\left\vert \psi \right\vert ^{2}$
disorder.

\subsubsection{Stability region of the RS solutions.}

The Hessian now has several additional terms

\begin{eqnarray}
H_{(ab)(cd)} &\equiv &\frac{n}{2}\frac{\delta ^{2}f}{\delta m_{ab}\delta
m_{cd}}=\left( m^{-2}\right) _{ac}\left( m^{-1}\right) _{db}+\left(
m^{-1}\right) _{ac}\left( m^{-2}\right) _{db}+4\delta _{abcd} \\
&&-q\left(
\begin{array}{c}
\frac{3}{2}\delta _{ac}\delta _{bd}Q_{ab}^{2}+\frac{1}{2}\left( \delta
_{ac}\delta _{bd}Q_{ba}^{2}+2\delta _{ad}\delta _{bc}Q_{ab}Q_{ba}\right) \\
2(\delta _{ac}\delta _{bd}Q_{aa}Q_{bb}+\delta _{bcd}Q_{ab}Q_{aa}+\delta
_{acd}Q_{ab}Q_{bb}) \\
+\delta _{ab}\left[ \delta _{cd}\left( Q_{ac}^{2}+Q_{ca}^{2}\right) +2\left(
\delta _{ac}Q_{ad}Q_{dd}+\delta _{ad}Q_{ca}Q_{cc}\right) \right]%
\end{array}%
\right)  \notag
\end{eqnarray}

For replica symmetric solutions $Q_{ab}=m_{ab}=\varepsilon \delta _{ab}+M$
the Hessian can be represented as

\begin{equation}
H=c_{+}I_{+}+c_{-}I_{-}+c_{U}U+c_{V}V+c_{J}J+c_{K}K+c_{N}N,
\label{operators}
\end{equation}%
where new operators $I_{\pm }$, $K,N$ are defined as
\begin{equation}
I_{\pm }\equiv \frac{1}{2}\left( \delta _{ac}\delta _{bd}\pm \delta
_{ad}\delta _{bc}\right) ;\text{ }K\equiv \delta _{ab}\delta _{cd};\text{ \
\ }N=\delta _{abc}+\delta _{abd}+\delta _{acd}+\delta _{bcd}.
\end{equation}
The coefficients are

\begin{eqnarray}
c_{+} &=&2\varepsilon ^{-3}-q\left( 3M^{2}+2\widetilde{m}^{2}\right) ; \\
c_{-} &=&2\varepsilon ^{-3}-q\left( M^{2}+2\widetilde{m}^{2}\right) ;  \notag
\\
c_{U} &=&4M^{2}\varepsilon ^{-5};c_{V}=-3M\varepsilon ^{-4};\text{ }  \notag
\\
\text{\ }c_{J} &=&4-q\left[ 5\left( \widetilde{m}^{2}-M^{2}\right) +8%
\widetilde{m}\left( \widetilde{m}-M\right) \right] ;  \notag \\
c_{K} &=&-2qM^{2};\text{ \ }c_{N}=-2q\widetilde{m}M.  \notag
\end{eqnarray}%
Generally the Hessian have four different eigenvalues: \cite{Alemeida}

\begin{equation}
\lambda ^{(1,2)}=c_{+}+\frac{1}{2}\left( c_{J}+4c_{N}\pm \sqrt{c_{J}\left(
c_{J}+8c_{V}+8c_{N}\right) }\right) ;\text{ \ }\lambda ^{(3)}=c_{+},\lambda
^{(4)}=c_{-}
\end{equation}%
Note that there are new matrices like $I_{+},I_{-}$ when $q\neq 0$. In the
case of $q=0$, $c_{+}=c_{-}$, so that only operator $I=I_{+}+I_{-}$ appears
in this case. Actually there is $\lambda ^{(4)}$ which is the eigenvalue of
Hessian on the antisymmetric matrices. However $\lambda ^{(4)}=c_{-}\geq 0$
is always hold on the RS solutions so that it can be ignored in determining
the instabilities of those RS solutions. Since the stability analysis is
quite complicated, we divide it into several stages of increasing complexity.

\subsubsection{\protect\bigskip Stability of the states on the diagonal -
off diagonal "transition" line}

The easiest way to see that the RS solutions can be unstable is to look
first at the transition line $a_{T}^{d}$, eq.(\ref{lowerl}). On the
transition line one has

\begin{equation}
c_{\pm }=c_{U}=c_{V}=c_{N}=0;\text{ \ \ }c_{J}=4-13q^{3/5};
\end{equation}%
and the eigenvalues simplify to%
\begin{equation}
\lambda ^{(3)}=0;\text{ \ \ \ }\lambda ^{(1,2)}=4-13q^{3/5}.
\end{equation}%
Therefore it is unstable for $q>q^{t}$
\begin{equation}
q^{t}=\left( \frac{4}{13}\right) ^{5/3},
\end{equation}%
marginally stable at a single point

\begin{equation}
a_{T}^{t}=-\frac{28}{13}\left( \frac{13}{4}\right) ^{1/3}\approx -3.2
\end{equation}%
and stable for $q<q^{t}$. We studied numerically the stability on both sides
of this line, see Fig.2. The diagonal (liquid) solution is stable below the
line ($a_{T}^{d}=2(3q^{2/5}-2q^{-1/5})$) for $q>q^{t}$. The line when $%
q>q^{t}$ the phase transition line (liquid to glass) is changed to a
different line which will be discussed in the next subsection.

\subsubsection{\protect\bigskip Stability of the diagonal solution}

Equation for $\widetilde{m}$, coefficients in Hessian and eigenvalues are:

\begin{equation}
-\widetilde{m}^{-2}+a_{T}+4\widetilde{m}-5q\widetilde{m}^{3}=0;
\end{equation}

\begin{eqnarray}
c_{+} &=&2\widetilde{m}^{-3}-2q\widetilde{m}^{2};\text{ \ \ }c_{J}=4-13q%
\widetilde{m}^{2}; \\
\lambda ^{(1,3)} &=&c_{+};\text{ \ \ }\lambda ^{(2)}=c_{+}+c_{J}=4+2%
\widetilde{m}^{-3}-15q\widetilde{m}^{2}.  \notag
\end{eqnarray}%
While $\lambda ^{(1)}$ is positive, $\lambda ^{(2)}$ is positive only below
the line defined parametrically via%
\begin{equation}
a_{T}=\frac{5-8\widetilde{m}^{3}}{3\widetilde{m}^{2}};\text{ \ \ \ }q=\frac{4%
\widetilde{m}^{3}+2}{15\widetilde{m}^{5}},  \label{upperl}
\end{equation}%
marked by dotted line in Fig.2. and this line is the phase transition line
(liquid to glass) when $q>q^{t}$. The former diagonal - off diagonal line
above the tricritical point is not a phase transition and is left as a light
dashed line to show that slope of the line below tricritical point and that
of the real transition line is different. It turns out that the line of eq.(%
\ref{upperl}) is a transition line into a RSB state, namely the
irreversibility line.

\subsubsection{Stability of the off diagonal solution}

The equations take a form:%
\begin{equation}
-\varepsilon ^{-2}+a_{T}+4\widetilde{m}-q\left[ 5\widetilde{m}%
^{3}-M^{3}-2M^{2}\widetilde{m}-2\widetilde{m}^{2}M\right] =0
\end{equation}

\begin{equation}
2\varepsilon ^{-3}-q\left( M^{2}+2\widetilde{m}^{2}\right) =0.
\end{equation}%
The coefficients in the expansion of Hessian are: Unlike the case of the $%
\left\vert \psi \right\vert ^{2}$ disorder, $\lambda ^{(1)}<0$ for each such
a solution. Therefore the diagonal state directly goes over into a RSB glass
state. It follows however two lines. The eq.(\ref{upperl}) above the
tricritical point and eq.(\ref{lowerl}) below it, see Fig. 2.

\section{RSB in the general case}

\subsection{General hierarchical gaussian variational Ansatz\protect\bigskip}

The free energy
\begin{equation}
\frac{f}{2n}=\sum_{a}\left\{ \left( m^{-1}\right) _{aa}+a_{T}m_{aa}+2\left(
m_{aa}\right) ^{2}\right\} -\sum_{a,b}\left\{ 2rm_{ab}^{2}-\frac{q}{4}%
(m_{ab}^{4}+4m_{aa}m_{bb}m_{ab}^{2})\right\}
\end{equation}%
lead on the replica symmetric sector to the following equations:%
\begin{equation}
-\varepsilon ^{-2}+a_{T}+4\widetilde{m}-2r\varepsilon -q\left[ 5\widetilde{m}%
^{3}-M^{3}-2M^{2}\widetilde{m}-2\widetilde{m}^{2}M\right] =0,
\end{equation}

\begin{equation*}
M\left[ 2\varepsilon ^{-3}-2r-q\left( M^{2}+2\widetilde{m}^{2}\right) \right]
=0.
\end{equation*}%
For the diagonal solution $M=0$ one obtains%
\begin{equation}
-\widetilde{m}^{-2}+a_{T}+4\widetilde{m}-2r\widetilde{m}-5q\widetilde{m}%
^{3}=0.  \label{qrdiageq}
\end{equation}%
The off diagonal solution on the bifurcation line obeys%
\begin{equation}
\widetilde{m}^{-3}-r-q\widetilde{m}^{2}=0.  \label{qroffdiageq}
\end{equation}

Hessian for the general RS solution takes a form of eq.(\ref{operators})
with coefficients
\begin{eqnarray}
c_{+} &=&2\varepsilon ^{-3}-2r-q\left( 3M^{2}+2\widetilde{m}^{2}\right) ;%
\text{ \ }c_{-}=2\varepsilon ^{-3}-2r-q\left( M^{2}+2\widetilde{m}%
^{2}\right) ; \\
c_{U} &=&4M^{2}\varepsilon ^{-5};\text{ \ \ \ }c_{V}=-3M\varepsilon ^{-4};%
\text{ \ \ }c_{K}=-2qM^{2};  \notag \\
c_{N} &=&-2q\widetilde{m}M;\text{ \ \ \ \ }c_{J}=4-q\left[ 5\left(
\widetilde{m}^{2}-M^{2}\right) +8\widetilde{m}\left( \widetilde{m}-M\right) %
\right] .  \notag
\end{eqnarray}%
On the bifurcation line it simplifies:%
\begin{equation}
c_{\pm }=2\widetilde{m}^{-3}-2r-2q\widetilde{m}%
^{2};c_{U}=c_{V}=c_{K}=c_{N}=0;c_{J}=4-13q\widetilde{m}^{2}.
\end{equation}%
The eigenvalues are

\begin{equation}
\lambda ^{\left( 1,2\right) }=c_{+};\text{ \ \ \ \ }\lambda
^{(3)}=c_{+}+c_{J}.
\end{equation}%
Therefore the Hessian vanishes $\lambda ^{\left( 1,2\right) }=\lambda
^{(3)}=0$ for the tricritical (branch) point defined by
\begin{equation}
r=\left( \frac{13q}{4}\right) ^{2/3}-\frac{4}{13}.  \label{tricritical}
\end{equation}

\subparagraph{Stability of the diagonal solution}

\bigskip Hessian and eigenvalues in this case are:

\begin{eqnarray}
c_{_{\pm }} &=&2\widetilde{m}^{-3}-2r-2q\widetilde{m}^{2};c_{J}=4-13q%
\widetilde{m}^{2};  \label{hessianrq} \\
\lambda ^{\left( 1,2\right) } &=&c_{\pm };\text{ \ \ \ }\lambda ^{\left(
3\right) }=c_{_{\pm }}+c_{J}=4+2\widetilde{m}^{-3}-2r-15q\widetilde{m}^{2}.
\notag
\end{eqnarray}%
Below the tricritical point we solve equation $\lambda ^{(1)}=0$
perturbatively in $q:$%
\begin{equation}
\widetilde{m}=r^{-1/3}\left( 1-\frac{q}{3}r^{5/3}\right) +O(q^{2})
\end{equation}%
and substitute $\widetilde{m}$ into eq.(\ref{qrdiageq}) to determine the
"weak disorder" part of the glass transition line:%
\begin{equation}
a_{T}^{g1}=-\frac{4-r}{r^{1/3}}+\left( \frac{5}{r}+\frac{4r^{4/3}}{3}\right)
q+O(q^{2}).
\end{equation}%
Above the tricritical point, namely for larger disorder, one solves the
equation $\lambda ^{(3)}=0$ perturbatively in $q$ around the tricritical
point of eq.(\ref{tricritical}), $q^{t}(r)=\frac{4}{13}\left( r+\frac{4}{13}%
\right) ^{3/2}:$

\begin{equation}
\widetilde{m}=\left( r+\frac{4}{13}\right) ^{-1/3}\left( 1-\frac{10}{24+13r}%
\ \Delta \right) +O(\Delta ^{2});\text{ \ \ \ \ \ }\Delta =\frac{q}{q^{t}(r)}%
-1.
\end{equation}%
The NEA RS solution is unstable everywhere as $c_{+}<0$ ($c_{+}<c_{-}=0$).
We therefore obtain the glass transition line with RSB in the general case.
To compare it with experiment one has to specify phenomenologically the
precise dependence of the GL model parameters on temperature. Next section
is devoted to this.

\section{Comparison of the theoretical and the experimental phase diagrams}

\subsection{General picture\protect\bigskip\ and comparison with other
theories}

As was discussed in Introduction, the interplay between disorder and thermal
fluctuations makes the phase diagram of high $T_{c}$ superconductors very
complicated. As a result of the present investigation together with the
preceding one \cite{Lidisorder} regarding the order - disorder transition
the following qualitative picture of the $T-H$ phase diagram of a 3D
superconductor at temperatures not very far from $H_{c2}(T)$ (so that the
LLL approximation is valid) emerges, see schematic diagram Fig. 3. There are
two independent transition lines.

\bigskip \textit{1. The positional order-disorder line.}

The unified first order universal order - disorder line comprising
the melting and the second peak sections separates the homogeneous
and the crystalline (Bragg) phases. The transition is therefore
defined by the translation and rotation symmetry only, and the
intensity of the first Bragg peak can be taken as order parameter.
The broken symmetry is not directly related to pinning, however
the location of the line is\textit{\ }sensitive to disorder. One
sees on Fig.3 that as the disorder strength $n$ increases the
melting line (solid line) curves down at lower point merging with
the second peak segment. The effect of disorder is quite minor in
the high temperature region, in which the thermal fluctuations
dominate, but become dominant at low temperatures. The line makes
a wiggle near the experimentally claimed critical point. The
\textquotedblleft critical point\textquotedblright\ is
reinterpreted \cite{Lidisorder} as a (noncritical) Kauzmann point
in which the latent heat vanishes and the line is parallel to the
$T$ axis in the low temperature region. The surprising
\textquotedblleft wiggle\textquotedblright , which appears in 3D
only, has actually been observed in some experiments.
\cite{Grover} It is located precisely in the region in which the
thermal fluctuations and the disorder compete. One might expect
that the line just has a maximum like in $BSCCO,$ but the
situation might be more complex. Thermal fluctuations, on the one
hand side, make the disorder less effective, the less disorder
effect will favor solid, but, on the other hand side, they
themselves melt the solid. The theoretical magnetization, the
entropy and the specific heat discontinuities at melting line
\cite{Lidisorder} compare well with experiments
\cite{Bouquet,Nishizaki}. The low temperature segment of the
disorder dominated second peak line is weakly temperature
dependent. Its field strongly decreases as the disorder strength
increases. \

\bigskip \textit{2. The glass transition line.}

The second line is the glass transition line, the dashed line in Fig.3. We
assume very small $q$ and draw on Fig.3 the glass lines for three different
values of $n.$ One observes as expected, that, as the disorder strength
increases, the line moves towards higher temperatures. We have not
calculated the glass line in the crystalline state, but anticipate that it
depends little on the crystalline order. This is consistent with
observations made in Ref. \onlinecite{Herbut} in which it was noticed (in a
bit different context of layered materials and columnar defects) that
lateral modulation makes a very small difference to the glass line although
it is obviously very important for the location of the order - disorder
line. So we just continue the glass line in the homogeneous phase into the
crystalline side (still marking it by a dotted lines). If the glass lines of
the liquid side and solid side join to a single glass line, then the glass
line must cross the ODT line right at the Kauzmann point. The theoretical
calculated intersection point in Fig. 3 marked by black blob do not appear
at the Kauzmann point. \ We attribute this to different approximations made
to draw the ODT\ and GT line. That the glass line crosses the ODT line right
at the Kauzmann point has been indeed supported by some evidences in an
experiment in $BSCCO$ \cite{Zeldov98}, thought two lines crossing exactly at
the Kauzmann point or not is still an open question experimentally and
theoretically.

Consequently there are four different phases, see Fig. 4: pinned solid (=
Bragg glass), pinned liquid (=vortex glass), unpinned solid (or simply
solid) and unpinned liquid (or simply liquid). The four phases of the vortex
matter are also expected to be present in the case of layered quasi 2D
superconductors and it was shown by in $BSCCO$ in \onlinecite{Zeldov98} that
the conductivity and the magnetization both indicate the glass line crossing
the ODT line near the maximum (Kauzmann point). There is evidence of the
crossing of the ODT and GT line also in $LaSCCO$ \cite{Forgan} and $YBCO$
\cite{Nishizaki,Taylor}, but the line seems to lie very close to the melting
line (the segment of ODT line in the high temperature). In Ref. %
\onlinecite{Taylor} a great care was paid to distinguish the two lines, and
it was found that at low field the GT line crosses again the melting line so
that the lower field segment of the melting line is again separating two
glassy or pinned states.

Now we compare the results on the phase diagram with other theories. The
only known theory providing both the ODT and the glass lines is the density
functional model. The picture advocated in Ref. \onlinecite{Menon02} on
basis of the replica density functional calculation in the framework of
thermodynamics of pinned line objects is qualitatively different from the
present one. In this theory the glass line does not intersect the ODT line
and therefore there is no unpinned solid phase. The comparison of the theory
studied in this paper with the replica density functional theory is
complicated by the fact that the applicability ranges of two theories are
different. In addition there are other phenomenological approaches, most
notably that of ref. \cite{Mikitik}, based on Lindermann criterion to map
the ODT line.

\subsection{Application to optimally doped YBCO\protect\bigskip}

We plot the ODT and the glass transition line in Fig. 5 along with the
experimental data on optimally doped $YBCO$ of Ref. \onlinecite{Nishizaki}.
The theoretical lines use the fitting parameters of \ ref. %
\onlinecite{Lidisorder} which fitted the ODT line and assume the $\left\vert
\psi \right\vert ^{4}$ disorder strength $q$ is small. We believe the value
of this parameter should be measured directly using a replica symmetry
breaking dynamical phenomenon rather than fitted in thermodynamics, though
it might be possible to adjust $q$, so that the curve can fit the glass
transition line better than the case of $q=0$. As was argued in %
\onlinecite{Lidisorder} (and found to be consistent with experiments on the
ODT line and the magnetization, entropy and specific heat discontinuities at
melting line) that the general dependence of the disorder strength on
temperature near $T_{c}$ is: $r(t)=n(1-t)^{2}/t$ with $n=0.12$. The formula
interpolates the one used at lower temperatures in ref. \onlinecite{Blatter}
with our dimensionless pinning parameter $n$ (proportional to the pinning
centers density) related to the \textquotedblleft pinning
strength\textquotedblright\ of Blatter $et$ $al$ \cite{Blatter} by $n=\gamma
\gamma _{T}^{0}/\pi Gi^{1/2}\xi ^{3}$. Note that experimentally the glass
transition line at lower fields is not measured precisely. Different
experiments locate it at various places using different criteria. As
mentioned above experimentally the GT line often crosses the melting line
again at very low fields. This question perhaps cannot be addressed within
the lowest Landau level approximation valid at fields above $1kG.$ The ODT
line in the melting region is very well established experimentally by great
variety of techniques. However the "second peak" segment is only poorly
determined due to difficulty to define it in the essentially dynamic
magnetization loops analysis. Recently developed muon spin rotation and
neutron scattering methods might be very helpful in that respect.

\bigskip

\section{\protect\bigskip Summary\protect\bigskip}

To summarize we considered the effects of both thermal fluctuations and
disorder in the framework of the GL approach using the replica formalism.
Flux line lattice in type II superconductors undergoes a transition\textbf{\
}into three \textquotedblleft disordered\textquotedblright\ phases: vortex
liquid (not pinned), homogeneous vortex glass (the pinned liquid or the
vortex glass) and the Bragg glass (pinned solid) due to both thermal
fluctuations and random quenched disorder. We show that the disordered
Ginzburg -- Landau model (valid not very far from $H_{c2}$) in which only
the coefficient of a term quadratic in order parameter $\psi $ is random,
first considered by Dorsey, Fisher and Huang, leads to a state with nonzero
Edwards - Anderson order parameter, but this state is still replica
symmetric. Namely there is no ergodicity breaking and no multiple time
scales in dynamics are expected. However, when the coefficient of the
quartic term in $\psi $ in the\ GL free energy also has a random component,
replica symmetry breaking effects appear (with ergodicity breaking). The
location of the glass transition line in 3D materials is determined and
compared to experiments. The line is clearly different from both the melting
line and the second peak line describing the translational and rotational
symmetry breaking at high and low temperatures respectively. The phase
diagram is therefore separated by two lines into four phases mentioned
above. In principle we could obtain the RSB solution near the phase
transition line by expanding the equations around the phase transition line
as in the spin glass theory, see, for example ref. \onlinecite{Parisi}, and
we found that the RSB is continuous. Thus RSB states involve multiple time
scales in relaxation phenomena.

It is natural to expect and is confirmed that the glass (irreversibility)
line crosses the \textquotedblleft order - disorder\textquotedblright\ line
not very far from its Kauzmann point. We are not sure if the crossing is
right at Kauzmann point. If two GT lines (on liquid side and solid side) are
joined, the crossing must be right at the Kauzmann point. We believe \
(speculate) that the glass line should cross the \textquotedblleft order -
disorder\textquotedblright\ line right at the Kauzmann point if the
experiments can be done accurately and the theory shall confirm it if the
model can be solved exactly. It is of great interest to solve a solvable toy
model to test this idea. The Kauzmann point is a point in which the
magnetization and the entropy difference between solid and liquid phases
changes sign. In this region the positive slope disorder dominated second
peak segment joins the thermal fluctuations dominated negative slope melting
segment. This is the region in which effects of the disorder and of the
thermal fluctuation are roughly of the same strength.

The replica symmetry breaking solution found here can be used to
calculate the detailed properties inside the glass state. This
however require generalization of the theory to include dynamics,
since most of irreversible phenomena are time dependent. In
particular it would be interesting to estimate the time scales
associated with quenched disorder. This is left for a future work.
Also we have considered only the three dimensional GL model here.
It can be applied to superconductors with rather small anisotropy.
It would be interesting to generalize the calculation to the
Lawrence - Doniach model and to the two dimensional case
describing thermal fluctuations and disorder in more anisotropic
layered superconductors and thin films..

\acknowledgments We are grateful to E.H. Brandt, B. Shapiro, A. Grover, E.
Andrei, F. Lin, J.-Y. Lin and G. Bel for numerous discussions, T. Nishizaki,
E. Zeldov and Y. Yeshurun, L. Paulus for providing details of their
experiments sometimes prior to publications The work was supported by NSC of
Taiwan grant The work of BR was supported by NSC of Taiwan grant
NSC\#93-2112-M-009-023 and the work of DL was supported the Ministry of
Science and Technology of China (G1999064602) and National Nature Science
Foundation (\#10274030). B.R. is very grateful for Weizmann Institute of
Science for warm hospitality during sabbatical leave.

\bigskip

{\Huge Figure captions}

{\LARGE Fig. 1}

Schematic $a_{T}-r$ plane phase diagram of the vortex liquid with the $%
\left\vert \psi \right\vert ^{2}$ disorder only. $a_{T}$ is the
LLL scaled temperature, while $r$ is the $\left\vert \psi
\right\vert ^{2}$ disorder strength. The dotted line in the is the
glass transition line. Below the line, the state is described by a
replica diagonal matrix $m_{ab}=m\delta _{ab}$, while above line
the vortex state has a nonzero Anderson Edwards parameter.

\bigskip {\LARGE Fig. 2}

Phase diagram with the $\left\vert \psi \right\vert ^{4}$ disorder
only in the $a_{T}$ $-$ $q$ plane. The dotted line marks the
replica symmetry breaking glass transition. The upper part above
the "tricritical" point of this line is given \ by
eq.(\ref{upperl}), while the lower part below the "tricritical"
point of the line is given by eq.(\ref{lowerl}) . The dashed line
is also given by eq.(\ref{lowerl}), but does not correspond to a
phase transition line. It is just a bifurcation line between two
replica symmetric (the diagonal and the off diagonal) states.

{\LARGE Fig. 3}

The $H-T$ Phase diagram for the $\left\vert \psi \right\vert ^{2}$ disorder
only ($q=0$) for three different disorder strength $n=0.12,$ $0.3$ and $0.08$%
. The dashed dotted line is$\ H_{c2}(T)$, the dotted line is the glass
transition line, the solid line is the order-disorder (ODT or liquid -solid)
transition line. Black blobs are the intersection points of the GT and the
ODT lines.

\bigskip {\LARGE Fig. 4}

The schematic phase diagram of $YBCO$. Four distinct thermodynamic
phases are: PL - pinned liquid (or vortex glass), PS - pinned
solid (or the Bragg glass), S - solid (unpinned solid), and L -
liquid (unpinned liquid).

\bigskip {\LARGE Fig. 5}

Comparison of the theoretical $YBCO$ phase diagrams with the experiment. \
The dotted and the solid lines are the theoretical glass transition and the
order--disorder transition lines respectively for the disorder strengths $%
n=0.12$ and $q=0$. $H_{m}(T),H_{sp}(T),H_{g}(T)$ are the experimental the
melting (ref.\onlinecite{Nishizaki}), the second peak (ref. %
\onlinecite{Bouquet} \cite{Bouquet}), and the glass transition (ref. %
\onlinecite{Nishizaki}) lines respectively.

\bigskip

\section{\protect\bigskip}

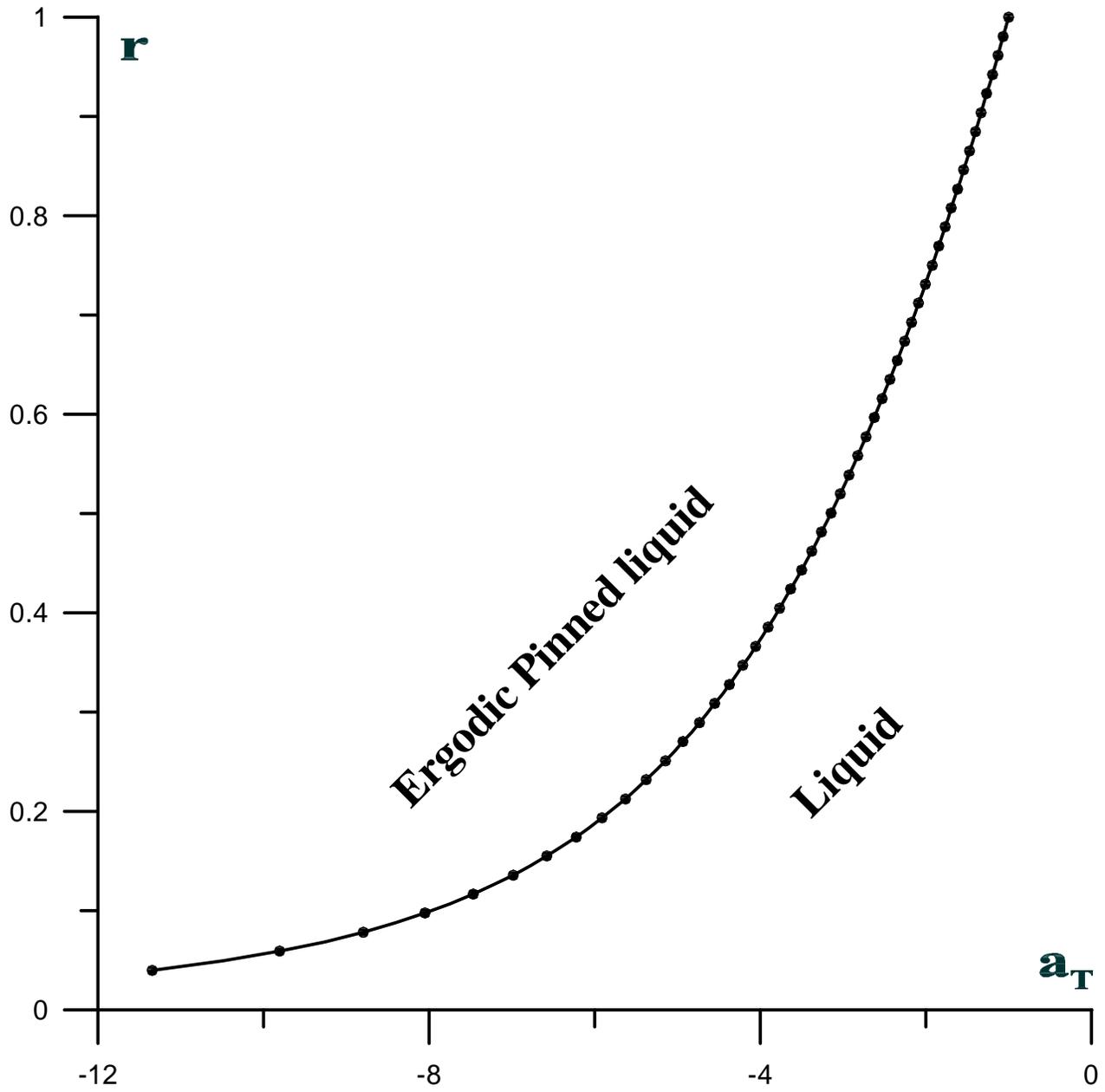
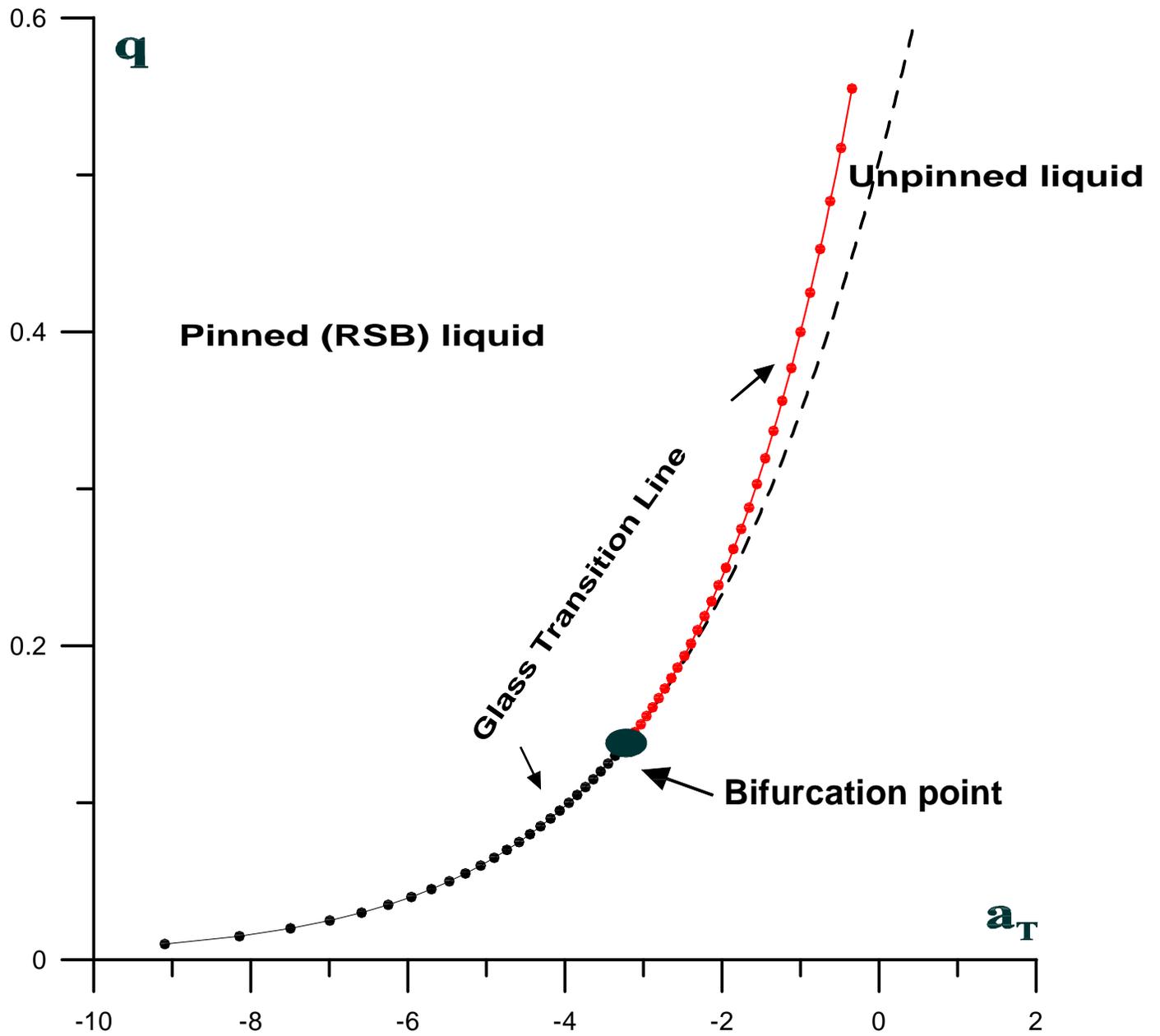
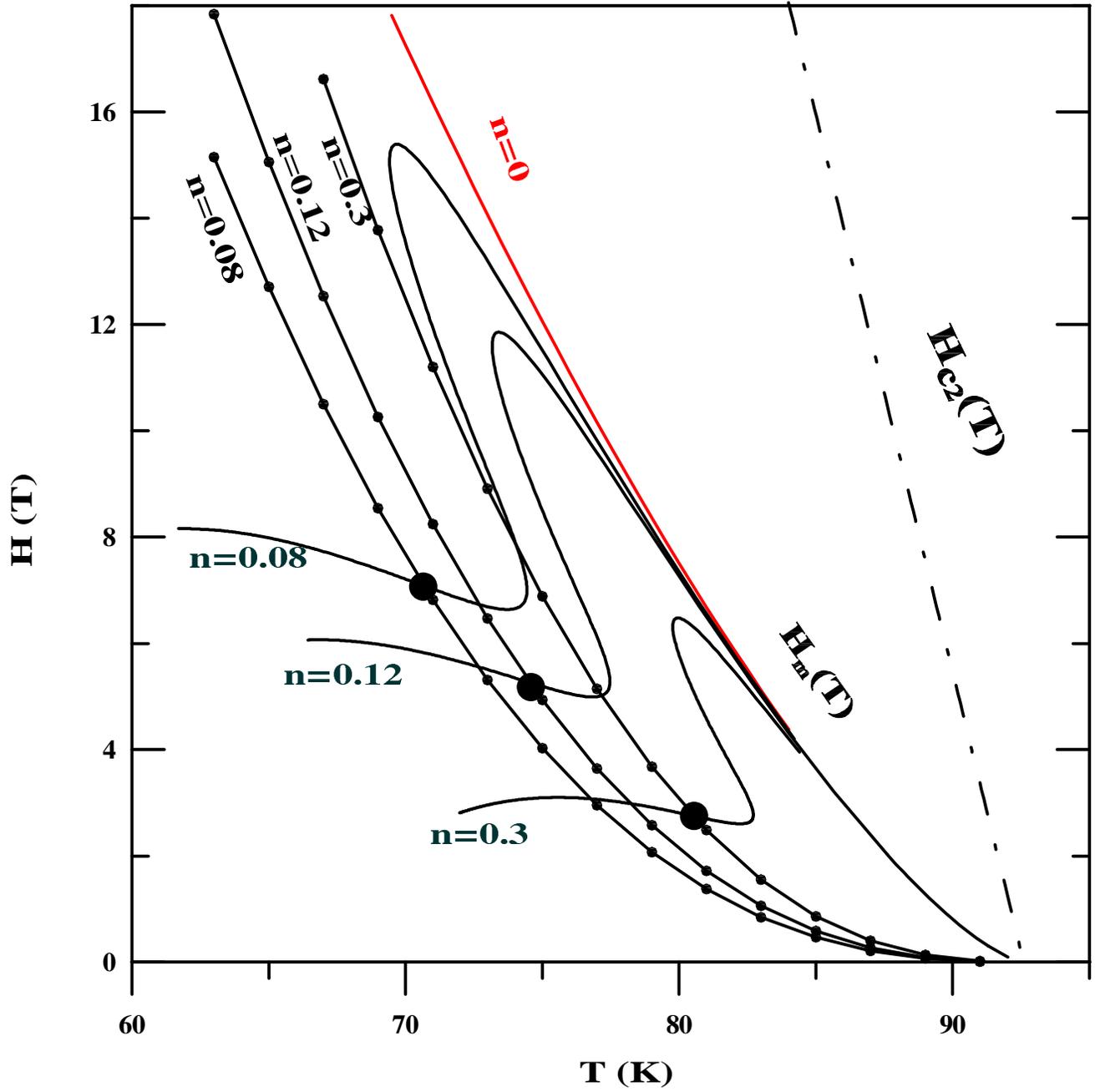
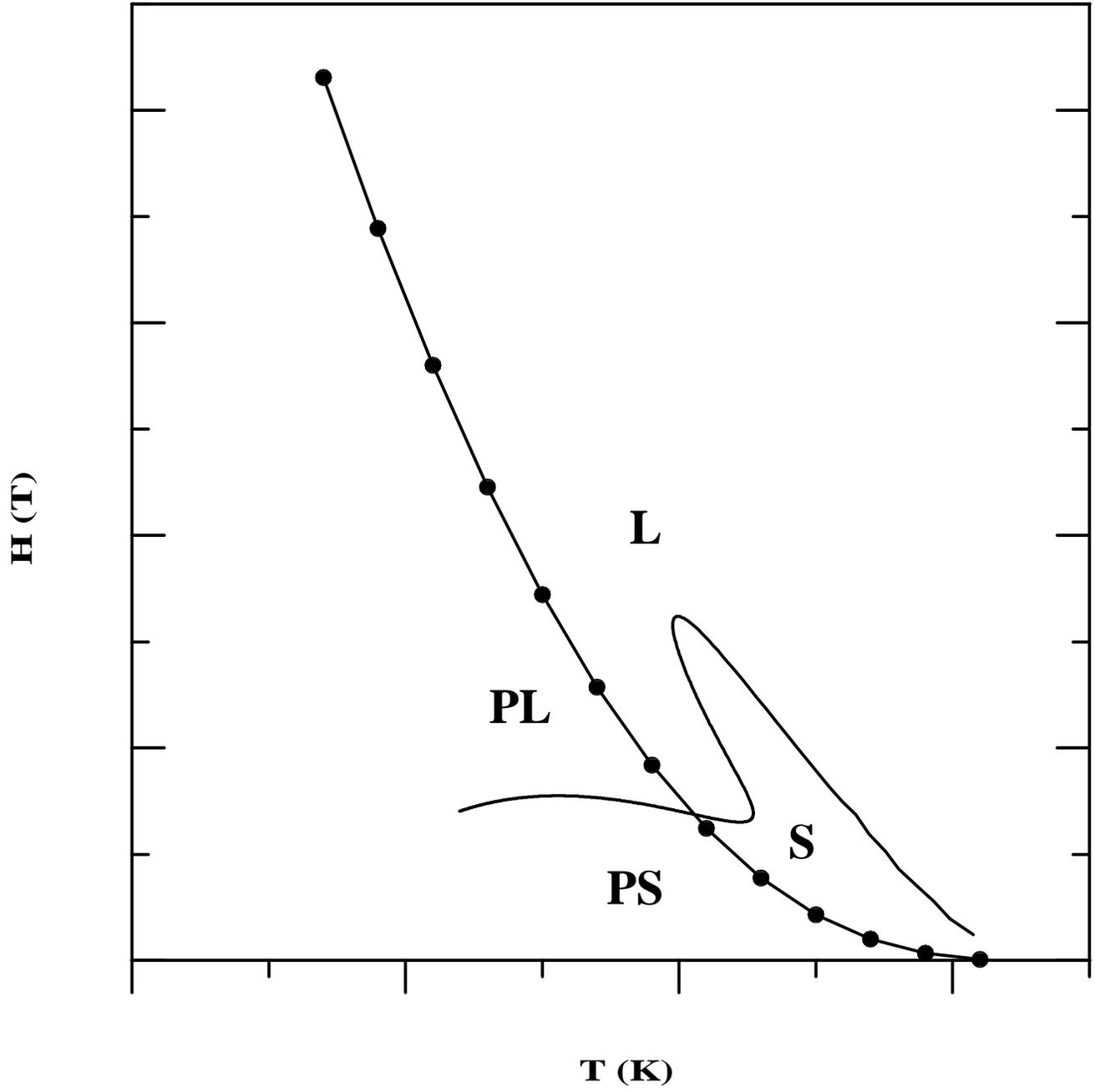
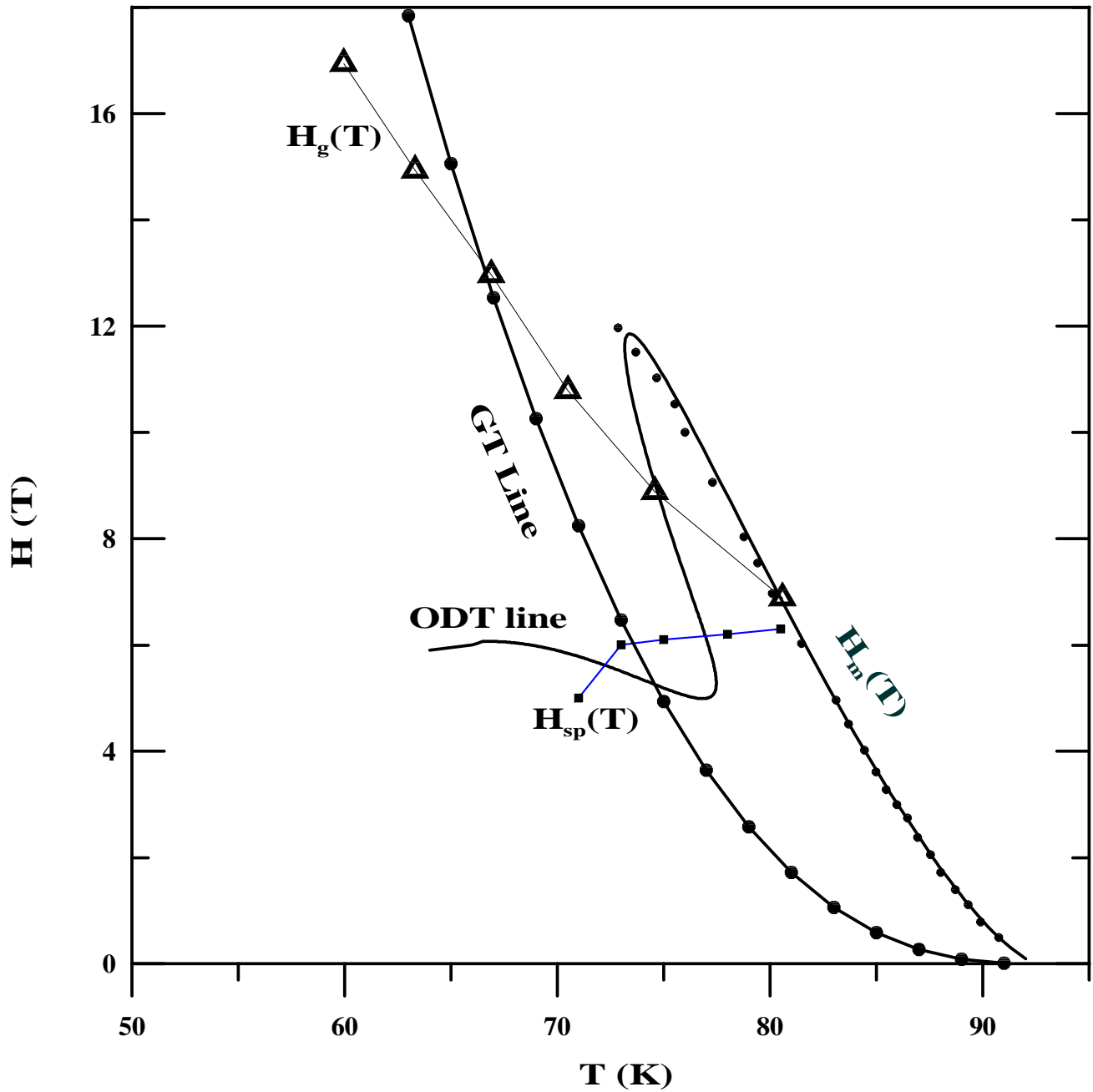
\end{document}